\documentclass[final,english]{bullsrsl}

\usepackage[latin1]{inputenc}

\usepackage[T1]{fontenc}

\usepackage{natbib} 

\usepackage{graphicx}

\begin{document}
\title{How to access astronomical observation facilities ?}



  \author[affil={}]{Micha\"el}{De Becker}

\affiliation[]{STAR Institute, University of Li\`ege, Belgium}



\correspondance{Michael.DeBecker@uliege.be}


\maketitle

\begin{abstract}
Access to astronomical data is a central component of astrophysical research. The allocation of telescope time is organized on an international scale through a highly competitive process. Over the past decades, this framework has evolved toward an increasingly professionalized system, particularly in the context of calls for telescope time proposals issued by major agencies or organizations, where hundreds of projects may compete for selection. Preparing a telescope time proposal is a demanding task for which junior researchers are not always adequately prepared. Astrophysicists typically acquire this expertise through first-hand experience, either by submitting their first proposal or by participating as members of a proposing team. At the same time, competition for telescope time is intense, and the significant effort invested in proposal preparation is accompanied by a non-negligible risk of rejection. This paper aims to present the general framework governing telescope time applications for both ground-based and space-borne observatories, with particular emphasis on the preparation of telescope time proposals. It discusses a set of key guidelines, some mandatory and others advisory, intended to help proposers navigate the application process more effectively, avoid common pitfalls and procedural missteps, and ultimately reduce the likelihood of preventable factors leading to a substantial decrease in the probability of selection. 
\end{abstract}





\section{Introduction}

Access to data is a key aspect of research activities in astrophysics. Whatever the main profile of astrophysicists -- observer or, to a large extent, theoretician -- their scientific results could not be achieved or validated without appropriate data gathered in well-selected spectral domains. A wealth of data is already available in multiple archive repositories throughout the world, most of which are open to the scientific community without restriction. Some of these data are locked for a given proprietary time after being collected at the request of a specific researcher, most of the time for one full year; beyond this limit, all data are accessible. Querying archives thus constitutes a valid way to obtain astronomical data of various types that may be used at any time by any interested researchers. 

Nevertheless, long-term archived and freshly obtained data result from the use of astronomical facilities at the request of users to pursue well identified scientific objectives. The operation of telescopes and their dedication to specific astrophysical topics and associated targets represent a huge pillar of modern astronomy. The path connecting the need for relevant data to their actual availability is not straightforward. Getting involved in the process of accessing astronomical facilities is demanding, time-consuming, and often a difficult task that requires some experience. 

What this paper focuses on is access to telescopes through the submission of {\it telescope time proposals}. The purpose of this paper is threefold.
\begin{enumerate}
    \item {\it Give an overview of the general context of the telescope time application process.} Knowing about the framework that leads to access to data through requests submitted to observatories is the bottom line of the process. Well-informed users are more prepared to participate, preventing them from making naive mistakes and allowing them to be more efficient in the planning of their applications. This topic is at the core of Sect.\,\ref{Sect_context}.
    \item {\it Clarify what a telescope time proposal is.} Working on a telescope time proposal is not an easy task, and it is important that any potential user knows very well what it is, what it contains, and how it will be processed after submission. This is addressed in Sect.\,\ref{Sect_proposals}.
    \item {\it Provide some guidelines to keep in mind when participating in the process of telescope time application.} When participating in the process, the preparation of telescope time proposals should comply with some rules, either formal or motivated by common sense. Complying with all the guidelines proposed in this paper may sometimes seem difficult, and neglecting some of them may not necessarily lead to a rejection of the project during the selection process. However, this can be seen as a list of recommendations intended to help proposers avoid some awkwardness or even mistakes and reduce the risk of falling in the ranking of competing projects during the evaluation procedure. This is the topic of Sect.\,\ref{Sect_guidelines}.
\end{enumerate}

\section{General context}\label{Sect_context}

Access to astronomical observation facilities is a wide topic that includes many specificities, which depend on the general observatory policy, the level of involvement of potential users in the instrumentation of interest in the observatory, or even potential funding participation to build or operate facilities. Many specific cases can be considered, but this paper focuses on the most general context that is of interest to any potential user, without concentrating on overly specific circumstances.

To simplify, one may distinguish between two main frameworks of access to astronomical facilities, which involve significantly different constraints and access policies:
\begin{enumerate}
\item {\it National / Institutional / Consortium observatories.} Some telescopes are limited in their access by the international community, simply because privileged access is guaranteed to a restricted scientific community. If the proposers are part of that community, they benefit from this privileged access with a low level of competition. This typically happens when an astronomical facility is operated by a unique institute, a consortium, or potentially by a national agency that restricts access based on national institute membership criteria. In such closed contexts, access to the telescope may be (almost) immediate or result from a low-competition selection managed locally (institutional to national level). In the most straightforward cases, the simple fact of being part of a consortium and knowing the people in charge of the facilities offers high guarantees that the data will be obtained, without any need to participate in a real competition. 
\item {\it Observatories open to the international community.} Most world-class observatories offer open observation time internationally, with no (or low) restrictions in terms of the proposers' institutional membership. Some of these facilities or agencies may be multi-national, such as the European Southern Observatory (ESO) or the European Space Agency (ESA). In addition, some observatories managed on a national scale are also open to the international community, such as the Giant Meterwave Radio Telescope (GMRT) in India or the National Aeronautics and Space Administration (NASA) and the National Radio Astronomy Observatory in the United States of America. In this framework, telescope time proposals must be submitted in response to a call for proposals (CfP, see Sect.\,\ref{Sect_proposals}), leading to a high competition level that significantly contributes to the difficulty of the exercise, especially for junior scientists not yet familiar with the process.
\end{enumerate}

The main focus of this paper is to describe the framework of {\it competitive access} to astronomical observation facilities that can represent a significant fraction of the working time of astrophysicists, especially at times of approaching deadlines of major calls for proposals. The competitive nature of the observation time application is a key aspect developed in Sect.\,\ref{Sect_compet}. 

\subsection{Competitive access}\label{Sect_compet}
In short, and before getting a bit deeper into the topic, the main idea prevailing here can be summarized by the following statement:

\noindent\makebox[\textwidth][c]{\it Many will apply, only a few will succeed!}

Within a framework of harsh competition, proposers must prepare for struggle,
\begin{enumerate}
\item {\it Against competitors.} Many smart people will also apply and propose relevant projects. Proposers should, from the beginning, renounce the thought that they are among the top tier competitors in the game. Most teams of proposers have already gained significant experience in preparing proposals, in particular, with that specific telescope. In addition, some projects will be resubmitted after being rejected in a previous call and will have benefited from feedback that helps the proposers significantly improve their project or at least mitigate some weaknesses pointed out by the reviewers (see Sect.\,\ref{Sect_resub}). Among the proposals submitted in response to a given call, typically there might be a very few percent of projects considered to be weak, and about the same proportion that may fall in the category of so-called excellent projects. All other projects, i.e. 80 to 90\,\% are very likely good or very good. When submitting a project, one should know that we are most probably part of this majority of projects that may deserve to be accepted, but only a few will be selected due to the high competition (see Sect.\,\ref{Sect_compet}).

\item {\it Against reviewers.} The people to whom the proposers are speaking are the reviewers (see Sect.\,\ref{Sect_how}). One should keep in mind that the people responsible for the evaluation will express a subjective opinion and should not expect any absolute objectivity from them. Evaluators are biased, depending on their own experience, their field of expertise, or simply their mood of the day. The task of evaluators is quite hard, as they have to propose a ranking and make a selection among many projects, most of which are at least good or very good. The situation is conducive to evaluators making a selection largely based on identifying a particular weakness, resulting in a devaluation of the project. As a result, some evaluators will search for reasons to reject a project rather than to support it. This also directly results from the high level of competition between projects that, on average, deserve some telescope time. Reviewers are thus the people to convince, and proposers must take this into account. This is the state-of-mind that is required to participate seriously in the competition.

\item {\it And more importantly... against themselves.} Reviewers are not the only protagonists being subjective, proposal writers are all deeply biased too. The assumption that evaluators' and proposers' biases will be well aligned is both reckless and unfounded. One of the most difficult aspects when preparing a proposal is trying to anticipate the criticism from reviewers, and above all, ensuring that what is written in the text is sufficient to carry the message and convince the readers. The preparation of a scientific justification (see Sect.\,\ref{Sect_content}) requires authors to step back from their own writing. It is easy to convince ourselves that our text is clear and effectively conveys our message. However, when reading our own text, we are more likely to revive the idea we had in mind while writing it rather than evaluate the text as it is. The proposers, thus, have to take some distance from their own writing and proceed with some kind of critical evaluation. Of course, a careful reading by collaborators who did not write the main part of the text is also highly recommended.
\end{enumerate}

As usual in scientific or technical domains, providing quantitative assessments is common practice. The measurement of the competitive nature of the telescope access is addressed in Sect.\,\ref{Sect_OF}.

\subsection{Measurement of the competition level}\label{Sect_OF}
Observatories make use of the concept of {\it oversubscription factor} to measure the level of competition for a given telescope resulting from a call for proposals. It is essentially defined as {\it the ratio of the cumulated requested time of all submitted projects over the total available telescope time during a given period}. In some cases, observatories distinguish between oversubscription (the ratio of accepted-to-submitted proposals) and pressure (the ratio of available-to-total requested time). In this document, the convention is to define the oversubscription factor in terms of time ratio, which aligns more closely with the policy of many observatories.

Some values for the oversubscription factor of various telescopes are provided in Table\,\ref{Table_OF}. The ranges of values account for the different results from various calls for proposals. One immediately notices that all values quoted here are greater than 1. This illustrates the fact that, in all cases, proposing teams altogether require more observation time than is actually available for the period of interest. {\it The higher the value, the harder the competition.} 

As a direct consequence of the high pressure on telescopes, 

\noindent\makebox[\textwidth][c]{\it Most proposals will be rejected, and the proposers  must be prepared to work}
\makebox[\textwidth][c]{\it on proposals with low expectation of obtaining the data}.

\begin{table}
\centering
\begin{minipage}{160mm}  
\centering
\caption{Ranges of oversubscription factors (OF) of a selection of ground-based and space-borne observatories. References : (a) \citet{Hainaut2022}; (b) \citet{Gemini}; (c) \citet{Iye2021}; (d) \citet{Keck}; (e) \citet{WHT}; (f) \citet{NRC2000}; (g) \citet{Gupta2017}; (h) \citet{GBT1,GBT2}; (i) \citet{Muxlow2012}; (j) \citet{STScI}; (k) \citet{HST2020}; (l) \citet{Parmar}; (m) \citet{Chandra}.}\label{Table_OF}
\end{minipage}
\begin{tabular}{lccclccclcc}
\hline
\multicolumn{3}{c}{Visible/infrared} & & \multicolumn{3}{c}{Radio} & & \multicolumn{3}{c}{Space telescopes} \\
\hline
Facility  & OF & Ref. &  & Facility & OF & Ref. &  & Facility & OF & Ref. \\
\hline
VLT & 2--5 & (a) & & VLA & 2--3.5 & (f) & & HST & 6--12 & (j,k)\\
Gemini & 1.5--3.5 & (b) & & GMRT & $>$\,2 & (g) & & XMM-Newton & 6--9.5 & (l)\\
Subaru & 3--5 & (c) & & VLBA & $\sim$\,2 & (f) & & Chandra & 4--6 & (m)\\
Keck & $\sim$\,5 & (d) & & GBT & 2--6 & (h) & & INTEGRAL & 20--4 (?) & (l)\\
WHT & 2--3 & (e) & & EVN & $\sim$\,2.2 & (i) & & JWST & 4.1--9 & (j)\\
\hline
\end{tabular}
\end{table}

Having a closer look at the numbers, one notices that for ground-based visible and infrared facilities, most large telescopes (several meters in diameter) are characterized by oversubscription factors reaching values of 5 to 7 for some calls for proposals. Recalling that 80 to 90\,\% of projects are good to very good, the rejection of several high quality projects is unavoidable. For major space observatories, it is even worse.

\subsection{Why do international observatories work according to a competitive approach?}

Observatories love high oversubscription, and they have some good reasons to do so.

\begin{enumerate}
\item {\it Measurement of the importance of the observatory.} A highly oversubscribed telescope is a facility that is used by an important scientific community, with many requests from potential users. This means that the telescope is efficient and provides a wealth of high quality data. This is consequently interpreted as a measurable indicator of the good quality of the observatory. The organization or agency operating the telescope can thus use this as justification for pursuing funding to support its operation or to upgrade the facility and further enhance its performance. A high oversubscription also opens doors for the funding of new or additional instruments to be mounted on the telescope.
\item {\it High level of science.} Given the high pressure on most modern telescopes, there is a fundamental need to make a selection. Although it is certainly difficult to objectively define discrimination criteria to select the so-called best projects, it is at least more feasible to identify the weakest projects and remove them from the competition. However, as already stated, weak projects represent only a few percent of the full amount of submitted proposals. The selection procedure also allows one to identify potentially excellent projects, but here again they represent at best only a few percent of the submissions.
\item {\it High cost of telescope building and operation.} The building of telescopes and their operation are expensive. The competitive framework of telescope time proposal selection is also supposed to warrant an appropriate use of the funding injected into these facilities. To give an idea of relevant orders of magnitude, the initial capital (including building) of several telescopes is provided in Table\,\ref{Table_TelCost}. In addition, the annual cost of full operation (including technical maintenance and operation, administrative costs, staff, etc.) typically represents several \% of the capital \citep{Goodrich2019}.
\end{enumerate}

\begin{table}
\centering
\begin{minipage}{160mm}  
\centering
\caption{Capital (expressed in equivalent 2019 million US\$) and starting year of operation for some astronomical facilities  \citep{Goodrich2019}.}\label{Table_TelCost}
\end{minipage}
\begin{tabular}{ccc}
\hline
Telescope & Capital & Year \\
\hline
VLA & 395 & 1976 \\
ALMA & 803 & 2004 \\
Keck I \& II & 405 & 1994 \\
Subaru & 350 & 2001 \\
LBT & 235 & 2008 \\
LSST & 660 & 2021 \\
\hline
\end{tabular}
\end{table}

\section{Telescope time proposals}\label{Sect_proposals}

\subsection{How does it work ?}\label{Sect_how}

\begin{figure}
\centering
\includegraphics[width=0.75\textwidth]{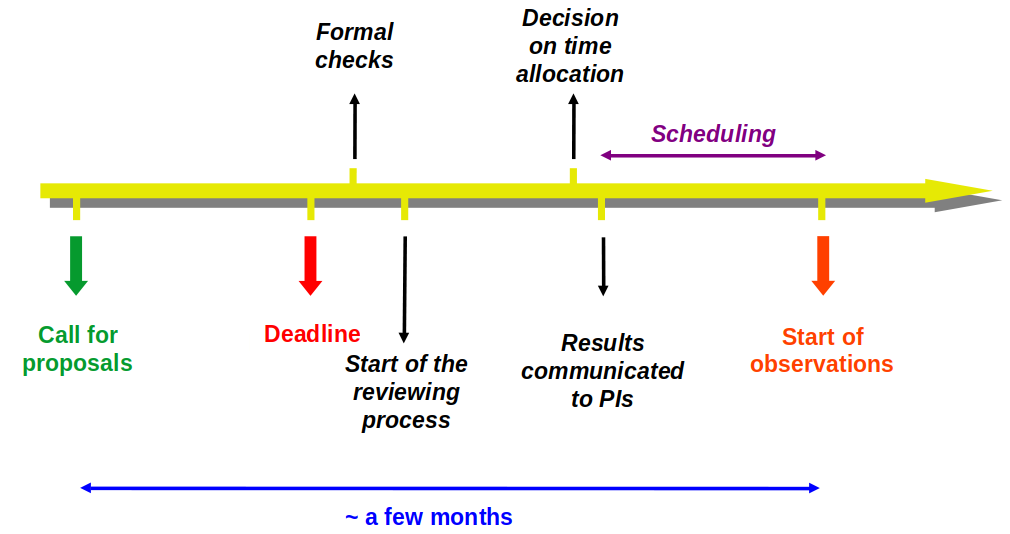}
\begin{minipage}{12cm}
\centering
\caption{Generic timeline of telescope time proposal processing, from the CfP to the start of observations.}\label{CfPTimeline}
\end{minipage}
\end{figure}

The typical timeline of the application procedure for telescope time is summarized in Fig.\,\ref{CfPTimeline}. The initial step is the {\it call for proposals (CfP)} or {\it announcement of opportunity (AO)}. This is the announcement made by the observatory or organization that manages access to a given telescope. These calls are periodic and occur most of the time once or twice a year (sometimes every 4 months). For space telescopes, there is usually one call per year \citep{Parmar,HSTCfP,JWSTCfP}. This is now also the case for access to ESO facilities \citep{ESOCfP} and for the MeerKAT radio observatory \citep{SARAOCfP}. Many ground-based observatories organize CfPs every 6 months, as in the case, for example, of the Devasthal Optical Telescope \citep{DOTCfP} and the GMRT \citep{NCRACfP}. Every CfP provides access to detailed instructions and guidelines on how to prepare the telescope time proposal with which the proposers have to be fully compliant (see Sect.\,\ref{Sect_content} and Sect.\,\ref{Sect_guidelines}).

Proposers must, in particular, be compliant with the {\it deadline} for proposal submission that is specified in every call. Nowadays, submissions are managed using on-line platforms that are automatically locked as soon as the deadline is reached. Proposing teams are highly encouraged to manage their time sufficiently in advance to deal with the telescope time proposal preparation, including communication with collaborators, to be ready on time. It is the responsibility of the Principal Investigator (PI), who is the spokesperson of the proposing team, to manage the proposal preparation and the submission procedure. The PI ensures that communication with co-investigators (coI) is smooth and that all coIs agree to the submission of the project as it is.

\subsection{Proposal reviewing}\label{Sect_rev}

All submitted proposals are received by the observatory or the organization operating the telescope and give rise to some formal checks. This is meant to verify the validity of the submission, notably in terms of the completeness of the proposal, before proceeding with the scientific and technical evaluation. 

Valid proposals are then managed by a committee that may be called {\it time allocation committee (TAC)} or {\it panel}. For every panel, a chairperson is appointed. The chair takes the responsibility of managing the discussions on evaluated projects and ensures that the output of the evaluation is compliant with the requirements of the observatory's policy. In short, two main approaches can be adopted for the scientific evaluation. On the one hand, the TAC, which is made up of several members (4 to 10, typically), proceeds with the scientific evaluation. All proposals are reviewed by all members who assign a mark (from 0 to 5 or from 0 to 10). In addition, every proposal is assigned to a primary and often a secondary reviewer. They are in charge of preparing a more comprehensive report on the proposals assigned to them. These reports are presented orally during a TAC meeting before being discussed by all members. These discussions lead to an adjustment of the ranking of the proposal, including some recommendations for time allocation. On the other hand, some panels do not proceed with the full scientific evaluation on their own. In that case, the TAC identifies potential external reviewers who will do the work. External reviewers are selected from the international scientific community. Most of the time, their names are extracted from a database held by the observatory, including PIs of previous calls for the same facility. The external reviewers can also simply be researchers identified by panel members without specific prior experience with the telescope of interest. The number of external reviewers contacted per proposal is highly variable, ranging between 3 and more than 10 in some cases. All external reports (including marks) are compiled by panel members and discussed during the panel meeting.

It is also common that, in parallel to the scientific evaluation, the observatory proceeds with a technical evaluation. This is typically performed by staff members of the observatory who are specialists in the use of the facility and who know the performance and limitations of the instrument.

At the end of the panel meeting, a full ranking of projects is achieved, with individual time allocation for proposals ranked above the cut-off defined by the total available time on that telescope. Most of the time, primary and secondary reviewers are in charge of preparing a brief feedback report that will be communicated to the PI only or to the full proposing team. As a result of this procedure, only a (low) fraction of proposals will be approved. In some cases, approved projects are assigned priorities (A, B, C). Typically, priority A projects are almost certain to be scheduled and executed. Priority B projects are very likely to be executed, but in the event of an unexpected issue, there is a risk that they may not be fully completed. Priority C projects are often considered fillers in the schedule. This means that after planning priority A and B projects in the schedule, empty slots can be filled with lower priority C projects. This works quite well for short duration projects without any time constraints, allowing some flexibility in the planning of the observations. The scheduling of observations is a difficult task that has to take into account many constraints, and it is managed by the observatory staff.

\subsection{The various categories of telescope time proposals}\label{Sect_categ}

Depending on the observatory, various kinds of proposal categories can be accommodated. The most common ones are briefly described below.

\begin{enumerate}
\item {\it Regular (or standard) proposals.} Most programmes submitted to observatories fit into this category. Such projects require low to moderate observation time: a few hours, a few nights, or a few orbits of a space telescope. 
\item {\it Large programmes.} Such projects are meant to address a more ambitious scientific question. In this case, the total observation time is significantly longer than for regular proposals. In some cases, observations under a large programme can extend beyond the observation period covered by the current CfP. For some observatories, very long observation times can be allocated to legacy projects (or key projects) aimed at producing long-lasting, high-value data sets. Such very ambitious projects are often conducted in close collaboration with the instrument team.
\item {\it Target of opportunity (TOO).} This consists of projects that require the triggering of observations at an unpredictable time and, therefore, cannot be included in the general schedule. A typical example is the investigation of transient sources, with no a priori idea of when such an event may occur. As a result, the proposal is tailored to get the opportunity to dedicate the telescope to the transient source as soon as the event is reported by another facility. That kind of flexibility is mandatory for studying unpredictable events. The triggered observation, therefore, interrupts the flow of scheduled observations, which requires a really strong science case. A strong requirement is the fast response of the observatory to trigger the observations before it is too late.
\item {\it Director discretionary time (DDT).} A small fraction of the total observation time is not directly offered in the framework of regular CfP. It is kept free in the event of exceptional requests made by a team of proposers that may not necessarily fit into the calendar and policy of regular calls. For instance, one may be dealing with a new idea that justifies the preparation of a future specific regular proposal, provided that some preliminary pilot project is organized beforehand to ascertain the feasibility of the final project. If the science case is strong enough, a DDT proposal may give access to a short observation that is used to validate the idea and make the final project stronger.
\item {\it Guaranteed time observation (GTO).} When a consortium develops a new instrument to be mounted on a telescope, or a country participates financially in its construction or operation, that consortium or country may be rewarded by the observatory through the so-called guaranteed observation time. Depending on the exact terms of the agreement between the consortium and the observatory, a given number of nights (or hours) per observation period may be dedicated to the members of the consortium. The allocation of the guaranteed time is not necessarily immediate and justifies the preparation of dedicated proposals that may be handled by the consortium itself, under the supervision of the observatory authorities.
\item {\it Science verification (SV).} When a new instrument is commissioned, the observatory organizes a science verification phase where members of the team that built the instrument proceed with some observations. However, in order to go further in the testing and involve the scientific community more directly, science verification can be open to any scientist willing to participate. SV programmes are typically quite short and aim to demonstrate the capabilities of a new instrument through insightful testing of its observing modes in support of relevant science cases
\end{enumerate}

\subsection{The typical content of proposals}\label{Sect_content}
A telescope time proposal must include some mandatory content presented in a way that is compliant with the templates provided by the observatory or the organization that manages the CfP. Although some differences appear from one case to another, certain pieces of content are systematically required for the proposal evaluation. This content may be presented as a unique document or spread over several dedicated ones.

\begin{enumerate}
\item {\it The cover sheet.} This is basically the front page of the full proposal. It includes at least the title, the abstract, the PI name and contact details, and the list of coIs with their affiliations and contact details.
\item {\it The scientific justification.} This is the main body of the proposal that is especially scrutinized during the review procedure. This includes a state-of-the-art overview of the scientific context, including an appropriate bibliography. The specific objectives of the project must be clarified. The proposing team must also provide some insight into the methodology that will be adopted, including a clear statement on how the data will be interpreted. The target(s) must also be identified and described. It is through the scientific justification that authors try to convince the reviewers of the relevance of their project.
\item {\it The observation sheet.} This is a more formatted document where the authors provide the details on the target(s), such as identifiers, coordinates, quantitative information (flux, magnitude, etc.), or any other relevant information. The exposure times and instrumental setup must be clarified. This part contains specific information that will be directly used by the technical staff of the observatory.
\item {\it The technical justification.} It consists of a description of the motivation for selecting a given instrumental setup in relation to the technical details provided by the documentation of that specific facility. Sometimes, it may also include comments for the telescope operators to complete the justification of the selected instrumental setup, depending on the specific needs for the submitted project.
\end{enumerate}

\section{Guidelines}\label{Sect_guidelines}

\subsection{The idea}\label{Sect_idea}

Having a clear idea of a scientific question to address is the primordial requirement to justify participation in the proposal submission process. Submitting a project for the sole purpose of participating is certainly not constructive. Serious participation necessitates a significant amount of work time, and dedicating it to a weak idea is a pure waste of precious resources.

\noindent\makebox[\textwidth][c]{\it This is the idea that should drive the need for a proposal, not the will }\\
\makebox[\textwidth][c]{\it to submit one that triggers the quick search for an idea.}

The initial idea is a necessary condition, but it is not enough. The idea must be developed with sufficient explanation to convince others that it is good and that the proposing team is about to do a great job with the data. Proposers must make sure they actually get a chance to compete before spreading their potentially excellent ideas. One should never forget that reviewers are also scientists. Although most scientists act with integrity, disseminating one's ideas always entails some risk.

\noindent\makebox[\textwidth][c]{\it It is advised not to share ideas without being prepared to defend them.}

In order to defend their ideas, the authors of the proposal have to be as pedagogical as possible. The content of the proposal must be self-consistent. Reviewers will not spend hours reading the bibliography provided in the scientific justification. Most likely, the evaluators are not specialists in the specific topic addressed in the proposal. The content of the proposal must be fine-tuned to be informative, clear, and well-justified. This is the price authors must be willing to pay if trying to get their ideas across is to be worthwhile.

\subsection{Who are you talking to?}\label{Sect_who}

\begin{figure}
\centering
\includegraphics[width=0.75\textwidth]{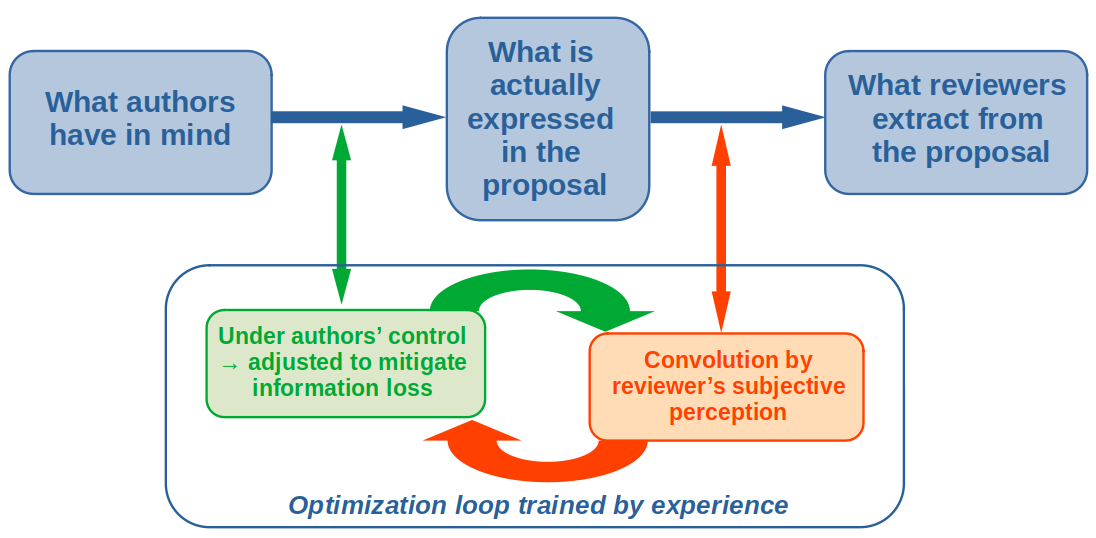}
\begin{minipage}{12cm}
\centering
\caption{Block diagram illustrating the information flow going from the authors' ideas and motivation to the perception by reviewers.}\label{BlockDiagram}
\end{minipage}
\end{figure}

As stated in Sect.\,\ref{Sect_compet}, the authors of proposals speak directly to the reviewers, who are human beings with their own sensitivities, strengths, and weaknesses. Figure\,\ref{BlockDiagram} illustrates the information flow and how it is altered from the thinking of the proposal to the perception by the reviewers. The authors convert what they have in mind into the text included in the proposal, which already alters the primary information. What reviewers get from it while reading is somewhat convolved by their own perception. Their scientific culture is at least reasonable and potentially excellent. However, most of the time, they are not specialists in the specific field of the proposal. Proposers should also remember that reviewers do not have time to explore the bibliography to gain a deeper understanding of the topic. They may have to evaluate many projects, and being part of the panel of evaluators is one mission among many other activities they have to deal with in their respective institutes. Participating in the evaluation is an important mission, but it adds to an already full agenda.

\makebox[\textwidth][c]{\it Authors should not expect reviewers to fill in the missing information for them.}\\

It is thus the authors' responsibility to prepare a project that efficiently translates the ideas, objectives, and methodology they have in mind. The better the writing, the less room there is for misinterpretation. In other words, the second blue arrow (on the right) in Fig.\,\ref{BlockDiagram} is less problematic if the first is well-adjusted. As a researcher gains experience in participating in this process, some kind of optimization loop is activated: benefiting from the feedback from reviewers (see Sect.\,\ref{Sect_resub}) allows proposers to anticipate some obvious mistakes or inaccuracies when working on future proposals, enabling them to improve the proposal writing and accordingly decrease the risk of misunderstanding by the reviewers.

As a last comment, reviewers and/or panel members are not stupid. Proposers are advised not to expect reviewers to believe everything they write in the scientific justification. This would very likely be interpreted as a bit offensive, with strong consequences for the proposal ranking. 

\noindent\makebox[\textwidth][c]{\it Upsetting reviewers is not a good approach to convince}\\
\makebox[\textwidth][c]{\it them about the relevance of a proposal.}\\

\subsection{Specific guidelines from the observatory}\label{Sect_obsguide}

Every CfP includes some documentation on the rules to follow to participate in the call. These guidelines are very important. A significant departure from these guidelines is likely to lead to the outright rejection of the proposal prior to any scientific evaluation. If the proposal is evaluated, not complying with the guidelines presents a very good opportunity for a severe downgrade of the project by the panel.

Some examples of typical guidelines are summarized below.
\begin{enumerate}
\item {\it Template.} In most cases, the observatory provides a template for the scientific justification. In such cases, authors must make use of it and comply with its structure. Such templates often include additional specific guidelines that the authors must adhere to.
\item {\it Page limit.} The size of the scientific justification is always limited. It can be restricted to 2, 3, or 4 pages, sometimes more when dealing with large or legacy programmes. Any size overflow will be treated as a mistake. Most authors make the effort to comply with the size restrictions, and considering oneself deserving of favorable treatment will not be well received by the panel. A page limit also applies when a separate technical justification is required.
\item {\it Font size.} Usually, a requirement is specified for the font size. Using tiny fonts to include as much information as possible to avoid page restrictions is not allowed. 
\item {\it Figures and tables.} Authors have to make sure they know the rules for including figures and tables in the scientific justification. Most of the time, these floating objects are included in the page limit. In that case, authors should not try to include them in addition to the allowed pages filled with text only. 
\end{enumerate}

\subsection{Be clear, be clear, be clear!}\label{Sect_clear}

In very short, a general rule applies:

\noindent\makebox[\textwidth][c]{\it Any lack of clarity will act against the proposal!}

Once again, we come back to the upper left part of Fig.\,\ref{BlockDiagram} (blue arrow on the left) about the need to make sure that the actual content of the proposal clearly and unambiguously reflects the authors' thoughts. The authors may have a clear idea of what they mean, but the reviewers may not. The authors have to be prepared to take a step back and reconsider what has been written.

The clarity of the proposal is also based on several specific points of attention summarized below.
\begin{enumerate}
\item {\it The abstract.} Often neglected, the abstract is an important component of the proposal. It may contribute to a non-negligible part of the reviewers' feelings. A well written abstract is not enough to save a poorly prepared scientific justification, but taking care of it can help support a well-prepared project.\\
\makebox[\textwidth][c]{\it A poor abstract is a missed opportunity to defend the project!}
When evaluators have already reviewed a given proposal, it is unlikely that they will read it again in depth several times before the panel meeting. However, it is very likely that they will return to some specific parts, focusing on key and well-emphasized pieces of information. In this context, the abstract represents a privileged area: it is short and should convey the core of the message. Consequently, neglecting its writing contradicts the requirement to drive the evaluators to a favorable state of mind.
\item {\it Layout and visual aspects.} What is included in a proposal is crucial, but how it looks is also very important. Within the boundaries and structure of the mandatory template, authors benefit from some flexibility in how to arrange the text. It is always better to read a well-structured text  with balanced paragraphs and indentations. When an evaluator opens a scientific justification and discovers overfull pages filled with text from the top left to the bottom right corner, this is never a good start. In the text, it is also recommended to make moderate use of italic and bold fonts. This is really helpful in highlighting some statements. It is an easy way to directly communicate with the readers, telling them that this highlighted information is a key point.
\item {\it Acronyms.} In any field of specialization, including research topics in astrophysics, the use of acronyms is very common. They constitute easy and straightforward ways to point to specific and well-defined concepts, with a minimum use of characters in a text. This makes it tempting to use many of them to save space in the limited area available for authors to develop their scientific justification. However, one should refrain from using them extensively. Too many acronyms may lead to a kind of "private joke" feeling for reviewers not familiar with their definitions. Even though acronyms are defined at their first use in the text, the evaluator may encounter most of them while reading the proposal, and this is not enough to make them familiar with such specific nomenclature, particularly if there are many. Moreover, the reading by reviewers will be interrupted at every occurrence, forcing them to go back in the text to return to the definition and consequently break the flow, at the expense of the readability of the project. Two or three acronyms are certainly more than enough.
\item{\it The objectives.} When describing the objectives, the text must be focused, punchy, and not too long. In a glance, the reader should clearly identify the purpose of the project. In this context, a common mistake found in many projects is the confusion between objectives and methodology. When this happens, the statement about the actual objectives is diluted in a description of the approach considered by the proposing team.

\noindent\makebox[\textwidth][c]{\it Objectives are about the "why", while the methodology is about the "how"!}

If the project is about obtaining images in a given spectral band of a specific field, the authors should not claim that "the objective is to obtain an image in that band". What must be clarified is why that image is needed : for instance, answering a scientific question, obtaining measurements to determine a physical quantity, etc. Once again, not being clear about the objectives is a missed opportunity to convince the evaluators.

\item{\it Direct language.} An important recommendation is to avoid approximate formulations. This gives the feeling of not being sure of what is claimed, and this is not reassuring for evaluators. The phrasing must be as direct as possible, avoiding overly long sentences. If evaluators have to read the same long sentences several times to understand the point, this is not in favour of the proposal.
\end{enumerate}

\subsection{The target(s)}\label{Sect_target}

Upon addressing the question of the target list that the proposal is focusing on, several aspects deserve to be kept in mind.

\begin{enumerate}
\item {\it The target selection.} In the case of any telescope time proposal, the target selection and description are key points. Authors must clarify their motivation for selecting that specific target or target list. If a target is selected from a class of objects, why is the project focusing on that one? What is special about this target? How does it fit into the broader context (the famous "big picture") of the science topic of interest? All these questions deserve to be addressed in some way by the proposing team. 
\item {\it The target description. } If some specific physical parameters regarding the target are required to better understand the relevance of its selection or to define the observation strategy, these parameters must be clarified in the scientific justification. For instance, if the target presents some variability, its variation time scale and its physical origin (if known) must be specified. Appropriate references should also be cited to support the validity of the statements included in the text.
\item {\it Quantitative assessment.} When asking for data about a given target using a given facility, it is always crucial to provide some quantitative information on the expected signal likely to be measured. Most observatories provide proposers with some on-line tools, such as exposure time calculators. Depending on the target and the type of astronomical data, a prediction on the brightness (or flux or flux density) may not be straightforward, but proposers are strongly encouraged to provide some motivated estimates or at least educated guesses on a valid range of values. In some cases, the proposing team may have access to models with appropriate prediction capabilities, leading to quantitative predictions that can be tested using the requested observations. In all cases, a loose quantitative prediction is much better than no estimate at all. The lack of quantitative assessment will more likely be considered a weakness or a high risk feature of the project: time allocation committees are often reluctant to grant observation time if no signal is likely to be measured, with no added value in non-detection.
\item {\it Existing data. } As a fundamental requirement, the authors must first check that the data they are asking for do not already exist. The proposal may not be needed if relevant data already exist in the archive of an observatory. Usually, panel members have access to a census of existing data associated with the targets of proposals they are evaluating, at least for the specific observatory of interest. 

\noindent\makebox[\textwidth][c]{\it Requesting data that already exist for a given target without}\\
\makebox[\textwidth][c]{\it appropriate justification is a direct cause of rejection!}

If additional data are needed, for example, to investigate the variability of the source, the need for the new data must be clarified in the proposal. Proposing teams are advised to retrieve existing data and show in the proposal what they could extract from it. It will also serve as evidence of the proposers' capability to deal with the data they are requesting.
\end{enumerate}

\subsection{The observatory/instrument/setup}\label{Sect_tech}

A proposal is always stronger when the motivation to use a specific telescope or instrument in a given setup is well presented. The authors must be very clear about the reasons why telescope time is requested for a given facility in a specific instrumental configuration. 

It is also very important that the authors are well aware of the advantages and limitations of the instrument and its selected setup. If the authors are aware of some restrictions imposed by the selected setup, it is advisable to anticipate and clarify (potentially in the technical justification) that this will not be a problem for the proposed project. This will prevent reviewers from pointing it out and potentially expressing concerns about it.

Some observatories offer high quality support for the preparation of proposals. Proposing teams should always consider contacting that support, especially in cases of low experience with that facility. Alternatively, having someone in the team who knows the technical aspects is always more than welcome. In some cases, it is even recommended by the organization managing the observatory to have a technical expert clearly identified in the team, with prior experience as PI or coI with the same observatory. This is, for instance, the case for the MeerKAT radio observatory in South Africa.

\subsection{Proposal resubmission}\label{Sect_resub}

It has been made clear in Sect.\,\ref{Sect_compet} that the competitive nature of the application for telescope time leads to many rejections. A rejection does not necessarily mean the project is definitely killed, as it may be resubmitted at a subsequent CfP. The important aspect of resubmission is to consider the content of the feedback report communicated by the panel after the previous submission.

Provided that the motivation for rejection was not a clear diagnostic of infeasibility, there is always room for improvement in a proposal. Although it can be somewhat frustrating to read the feedback report, the most constructive attitude is to read and analyze the comments with care. Even when comments do not seem very clear to the authors, at least the point raised by the reviewers indicates that something could have been made clearer in the proposal content.

\noindent\makebox[\textwidth][c]{\it It is always better to revise a proposal, taking inspiration from}\\
\makebox[\textwidth][c]{\it vague comments rather than completely ignoring them!}

Not taking into account comments from a previous submission will be considered a mistake, and the project will be severely downgraded. Panel members will consider that they are wasting their time reviewing and commenting on projects. This can even be a straightforward reason for an easy rejection. When resubmitting a proposal, the authors should keep in mind that some panel members or external reviewers active in the previous CfP will still participate in the evaluation during the current one.

Finally, although ignoring previous comments is most probably leading to a rejection, taking them into account does not offer any guarantee of success. One has to remember that the reference frame for the evaluation of the revised proposal is not limited to the feedback report. It encompasses the full series of competing proposals submitted in the new CfP along with a panel of reviewers that may have been slightly (or more deeply) modified. A resubmission is not evaluated separately from the full amount of proposals, and, among other projects, several will also be resubmissions from previous calls. The revision of a project along the lines of a feedback report is a necessary condition to have a better chance of success, but it is not necessarily sufficient to get the telescope time at the next attempt.

\subsection{Being in a hurry!}\label{Sect_rush}

\begin{figure}
\centering
\includegraphics[width=0.75\textwidth]{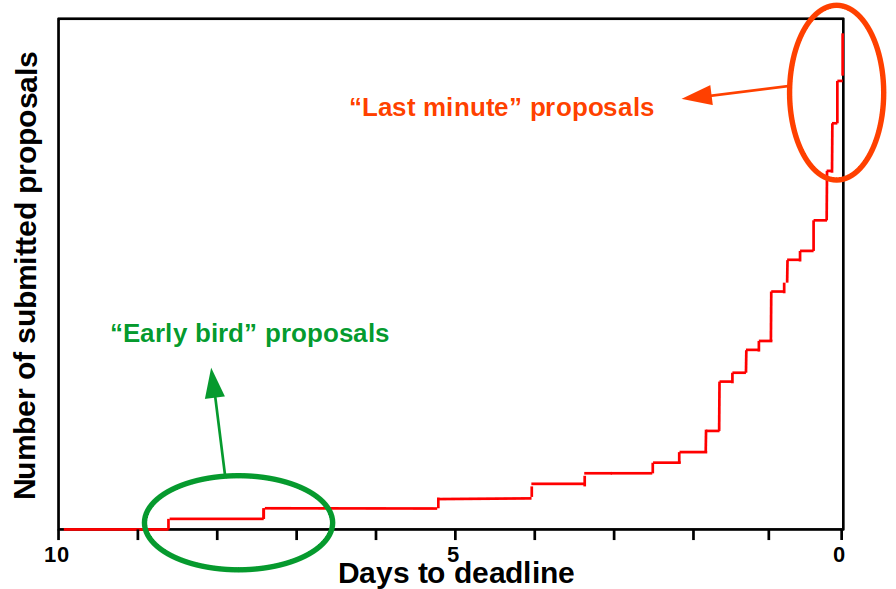}
\begin{minipage}{12cm}
\centering
\caption{Typical shape of a proposal submission curve, inspired by the the results valid for several observatories. It illustrates the growth of the cumulated number of submitted proposals as a function of remaining time to the deadline.}\label{SubmissionCurve}
\end{minipage}
\end{figure}

A key date in the timeline represented in Fig.\,\ref{CfPTimeline} is the deadline for submission. That date is clearly communicated in advance, allowing the proposing team to work ahead of that deadline to prepare mature and well-written projects. However, in practice, scientists are always over-busy with multiple commitments, which frequently pushes astrophysicists to feel rushed in the preparation of the project, or at least in its finalization. Consequently, most projects are submitted on the last day, with a significant fraction submitted even during the last hours before the deadline. This is illustrated in Fig.\,\ref{SubmissionCurve}, where the cumulative number of submitted proposals is plotted as a function of days to the deadline. Although that plot is not a real one measured for a specific observatory, it is quite typical of the situation encountered in most observatories. The preparation of the proposals submitted in the last couple of hours did not necessarily start late, but the authors (or the PI only) may have made the choice to fine-tune the project until the last minute. The writing of some late projects may have started several days in advance, but several reasons can lead to a late submission (the availability of coIs, issues in processing a preliminary data set, the perfectionism of the PI, etc.). 

As far as timing is concerned, some recommendations could be formulated.
\begin{enumerate}
\item {\it Availability of tools.}  The preparation of the project may require preliminary modelling that necessitates the availability of ready-to-run software. If you are in a hurry, this is not the right time to realize that the installation of these tools crashes.
\item {\it Unexpected issues.} A hard drive crash ? A personal commitment at home? The PI gets sick two days before the deadline? There are many unpredictable circumstances that can substantially reduce the availability of PIs and coIs at the wrong time. Anticipating the deadline is very likely to save a project by mitigating the issue of last minute unavailability.
\item {\it Internet connection.}  It is highly recommended not to rely too much on the stability of the internet connection during the last minutes before the deadline. As submissions proceed through on-line platforms, an internet issue may compromise all authors' efforts at the doorstep of the submission portal. 
\end{enumerate}

\section{Concluding remarks}\label{Sect_concl}

The application for telescope time is a highly competitive process that is intrinsically part of the work of astrophysicists. Accessing data is a demanding task that can be somewhat frustrating, but it is clearly worth the effort. All successful proposers have tried ! Some of their projects were very likely rejected before succeeding in the next attempt after a constructive review of their work. When a project is rejected, the best attitude consists of seeing this as an opportunity to come back to it and to grow as a scientist.

\noindent\makebox[\textwidth][c]{\it You either win or you learn !}

Principal investigators, and in particular young people new to this practice, have to remember that most projects are not approved. This is obvious when looking at the oversubscription factors of most observatories. Not being part of the lucky winners does not mean that the authors are not good professionals. It simply means that the level of competition is high enough to prevent even good scientists from getting what they asked for.

\noindent\makebox[\textwidth][c]{\it What if it did not work ? ...  No shame ! Keep pushing !}

As a last comment, let us emphasize that one day proposers may become evaluators. Participating in this process from the other side is the perfect complement to being a proposer. Being an evaluator, especially as part of a time allocation committee, is also highly demanding. It is hard work that takes time and requires a lot of effort, especially for new participants. However, this is also a great opportunity to learn a lot. Reviewers occupy a privileged position not because of their "power to support or kill projects" (which would reflect an inappropriate motivation) but especially because they have access to an overview of the various ways good scientists defend projects and ideas. This is thus a great opportunity to learn how to become more efficient in preparing telescope time proposals. Besides, it is, of course, a wonderful occasion to broaden one's scientific culture. As clarified in this paper, authors have to be clear and pedagogical. Having the opportunity to access such a diversity of short writings on various topics that reviewers are not necessarily familiar with allows them to enrich their background. Every astrophysicist should be encouraged to participate for these excellent reasons.


\begin{acknowledgments}
The author warmly thanks the organizers of the workshop for this opportunity to present this topic as an invited lecture. A big {\it "thank you"} also goes to the participants of the meeting, both for the pleasant atmosphere during the event and for the enthusiastic feedback received after giving this talk.

\end{acknowledgments}

\begin{furtherinformation}

\begin{orcids}

  \orcid{0000-0002-1303-6534}{Micha\"el}{De Becker}
  
\end{orcids}



\begin{conflictsofinterest}
The author declares no conflict of interest.
\end{conflictsofinterest}

\end{furtherinformation}



%

\end{document}